\documentstyle[pra,floats,eqsecnum,aps]{revtex}
\begin{document}

\title{Four-potentials and Maxwell Field Tensors from $SL(2,C)$
Spinors as Infinite-Momentum/Zero-Mass Limits of their Massive
Counterparts}

\author{S. Ba{\c s}kal~\footnote{On leave from Department of Physics,
Middle East Technical University, 06531 Ankara, Turkey} and Y. S. Kim\\
\small{\it{Department of Physics, University of Maryland, College Park,
Maryland 20742, U.S.A.}}}

\maketitle

\begin{abstract}
Four $SL(2,C)$ spinors are considered within the framework of
Wigner's little groups which dictate internal space-time
symmetries of relativistic particles.  It is indicated that the
little group for a massive particle at rest is $O(3)$, while it is
$O(3)$-like for moving massive particles.  The little group becomes
like $E(2)$ in the infinite-momentum/zero-mass limit.  Spin-$\frac{1}{2}$
particles are studied in detail, and the origin of the gauge degrees
of freedom for massless particles is clarified.  There are sixteen
different combinations of direct products of two $SL(2,C)$ spinors
for spin-1 and spin-0 particles.  The state vectors for the $O(3)$
and $O(3)$-like little groups are constructed.  It is shown that in the
infinite-momentum/zero-mass limit, these state vectors become scalars,
four-potentials and the Maxwell field tensor.   It is revealed that
the Maxwell field tensor so obtained corresponds to some of the state
vectors constructed by Weinberg in 1964 [S. Weinberg, Phys. Rev.
{\bf B135} 1049 (1964)].\\

\end{abstract}

\def\a{\alpha}
\def\ad{\dot \alpha}
\def\b{\beta}
\def\bd{\dot \beta}
\def\epe{e^{\eta}}
\def\eme{e^{-\eta}}

\section{Introduction}\label{intro}

One of the beauties of Einstein's special relativity is the unified
description of the energy-momentum relation for massive and massless
particles through $ E=[(cP)^{2}+(Mc^{2})^{2}]^{1/2}$.  In 1939 Wigner
observed that in addition to mass, energy and momentum, relativistic
particles have internal space-time degrees of freedom~\cite{wig39}.
For this purpose, he studied the subgroups of the Poincar\'e group whose
transformations leave invariant the four-momentum of a given free particle.
The Poincar\'e group is the group of inhomogeneous Lorentz transformations,
namely Lorentz transformations followed by space-time translations.  The
two Casimir operators of this group correspond to the (mass)$^{2}$ and
(spin)$^{2}$ of a given particle.  The maximal subgroup of the Lorentz
group which leaves the four-momentum invariant is called the little group.
Wigner showed that the internal space-time symmetries of massive and massless
particles are dictated by the $O(3)$-like and $E(2)$-like little groups
respectively.  Here, the word ``like'' is used to indicate that the Lie
algebra is the same.  The $O(3)$-like little group has the same Lie algebra
as that of $O(3)$.

Electron spins are manifestations of internal space-time structure of
elementary particles.  They are described by the Dirac equation, which
constitutes a non-unitary finite-dimensional representation of Wigner's
little group.  Indeed, the Dirac equation is a four-by-four representation
of the $SL(2,C)$ group.  In this paper, we shall examine systematically
the $SL(2,C)$ spinors.  While there are two orthogonal spinors in the Pauli
representation of SU(2), there are four different spinors in the $SL(2,C)$
regime~\cite{berest89,knp86}.  From these, we shall construct
representations of the $O(3)$-like little group, $E(2)$-like little group,
and the contraction of the $O(3)$-like little group to the $E(2)$-like
little group in the infinite-momentum/zero-mass limit.

It was shown by In\"on\"u and Wigner that the
rotation group $O(3)$ can be contracted to $E(2)$ ~\cite{inonu53}.
An important development along this line of research is the application
of group contractions to the unification of the two
different little groups for massive and massless particles.
The $E(2)$-like little group for massless particles is obtained
from the $O(3)$-like little group for massive particles in the
infinite-momentum/zero-mass limit~\cite{misra76}.

Like the three-dimensional rotation group, $E(2)$ is a
three-parameter group.  It contains two translational degrees
of freedom in addition to the rotation.  The physics
associated with the translational-like degrees of freedom for
the case of the $E(2)$-like little group has been shown by various
authors to be the gauge degrees of freedom for massless
particles~\cite{janner71,hks85}.  Indeed, these gauge degrees of freedom
emanate from the contraction of the transverse components of the rotation
generators during the contraction process of the $O(3)$-like little
group to the $E(2)$-like little group~\cite{hks83}.  This contraction
process has been studied in detail by Kim and Wigner in terms of
cylindrical geometry~\cite{kiwi87jm}.

As far as $SL(2,C)$ is concerned, it has been possible to interpret
the Dirac spinors within the framework of Wigner's representation
theory~\cite{knp86}.  Is it then possible to find a place for Maxwell's
field tensors in the same theory?  In 1964, Weinberg formulated the
process of constructing all representations for massless fields starting
from the $SL(2,C)$ spinors~\cite{wein64}.  However, he uses only the
spinors which are invariant under the translation-like transformations
of the $E(2)$-like little group.  Since these transformations are
gauge transformations, Weinberg's construction contains the
gauge-invariant Maxwell field tensor.  Weinberg's construction does
not include gauge-dependent four-potentials.

It has earlier been discussed in the literature that the electromagnetic
four-potential can be obtained from the group-contraction
procedure~\cite{kiwi87jm}.  But the contraction procedure for the field
tensor has not been discussed.  We address this problem in the present
paper.  In order to have a coherent presentation of this procedure, we
construct both the four-vector and the second-rank tensor from the four
$SL(2,C)$ spinors for massive particles at rest.  We then boost those
spinors resulting in boosting the four-vector and the second-rank
tensor.  When the boost parameter approaches infinity, they become
the electromagnetic four-potential and the Maxwell field tensor.  In
this way we establish a covariant picture of the little group and its
representation space for both massive and massless particles.

In Sec. \ref{sec2}, we review the little groups and emphasize that they
are Lorentz-covariant entities.  The $O(3)$-like little group is not
only valid in the rest frame of the massive particle, but also in all
other Lorentz frames.  If the particle is not at rest, the little group
is not $O(3)$, but is an $O(3)$-like group which is a covariant entity.
In Sec. \ref{spinors}, the spin 0 and spin 1 representations of the
little group are constructed from the $S(2,C)$ spinors.  Since there are
four $SL(2,C)$ spinors for a spin-$\frac{1}{2}$ particle, there are
four different types of combinations of the usual $SU(2)$ construction.
In Sec.~\ref{4vec}, four-vectors and second rank tensors are constructed
from the $SL(2,C)$ spinors.  They are covariant entities which can be
boosted from the rest frame to an infinite-momentum frame.  Within the
framework of the In\"on\"u-Wigner contraction, they become the
four-potentials
and the Maxwell field tensor.  It is noted that the field tensor
corresponds
to some of the state vectors constructed by Weinberg.

\section{Little Groups of the Poincar\'e Group}\label{sec2}
The group of Lorentz transformations is generated by three
rotation generators $J_{i}$ and three boost generators $K_{i}$.
They satisfy the commutation relations of the $SL(2,C)$ Lie algebra~:
\begin{equation}
\begin{array}{rrr}
[J_{i},J_{j}]=i \epsilon_{ijk} J_{k}, \quad
[J_{i},K_{j}]=i \epsilon_{ijk}K_{k}, \quad
[K_{i},K_{j}]=-i \epsilon_{ijk} J_{k}.
\end{array}   \label{com1}
\end{equation}
For a massive point particle there is a Lorentz frame in
which the particle is at rest.  In this frame, the little
group is clearly the three-dimensional rotation group $O(3)$.
The four-momentum is not affected by this
rotation, but the spin variable changes its direction.
Hence, the little group of the moving massive particle
can be obtained by boosting along the direction of the
momentum.  Without loss of generality, if the particle is
boosted in the $z$ direction, the generators of
the little group can be obtained by
\begin{equation}
J'_{i}=B(\eta)J_{i}B(\eta)^{-1} ,
\end{equation}
where $B(\eta)=\exp(-i\eta K_{3})$.  Since $J_{3}$ commutes
with $K_{3}$, it remains invariant, while $J_{1}$ and
$J_{2}$ assume the form
\begin{equation}
J'_{1}=\cosh \eta J_{1} + \sinh \eta K_{2} , \qquad
J'_{2}=\cosh \eta J_{2} - \sinh \eta K_{2} ,
\end{equation}
which satisfy the Lie algebra of $O(3)$.  Thus, a moving massive
particle still has its $O(3)$-like little group.

However, for massless particles there are no Lorentz
frames in which the particle is at rest.  The approach is to
consider the limiting case in which the mass of the
particle becomes vanishingly small yielding the
boost parameter to become infinite.
After renormalizing the generators $J'_{1}$ and $J'_{2}$ as
\begin{equation}
N_{1}=-(\cosh \eta)^{-1} J'_{2},  \qquad
N_{2}=(\cosh \eta)^{-1} J'_{1}
\end{equation}
in the infinite-$\eta$ limit they reduce to
\begin{equation}\label{n1n2}
N_{1}= K_{1} - J_{2}, \qquad N_{1}= K_{2} + J_{1}.
\label{com2}
\end{equation}
The operators $N_{1}, ~N_{2}$ and $J_{3}$ satisfy
the commutation relations
\begin{equation}
[J_{3},N_{1}]=i N_{2}, \qquad
[J_{3},N_{2}]=-i N_{1}, \qquad
[N_{1},N_{2}]= 0 ,
\end{equation}
where $J_{3}$ is like the rotation generator, while
$N_{1}$ and $N_{3}$ are like translation generators
in the two-dimensional Euclidean plane.  Hence,
they are the generators of the $E(2)$-like little group
for massless particles.

The traditional approach to the little groups has been to emphasize the
difference between those for massive and massless particles.  In this
paper, we would like to emphasize that the little group is a covariant
entity and remains $O(3)$-like in all Lorentz frames.  It becomes $E(2)$-like
in the infinite-momentum/zero-mass limit as in the case of the
In\"on\"u-Wigner contraction in which $O(3)$ becomes $E(2)$.

\section{$SL(2,C)$ Spinors}\label{spinors}
The commutation relations of the three dimensional rotation group
is contained in the $SL(2,C)$ algebra introduced in (\ref{com1}) as
\begin{equation}
[J_{i},J_{j}]=i \epsilon_{ijk} J_{k}.
\end{equation}
The two-by-two representation of this group is called $SU(2)$, and the
generators are given in terms of the Pauli spin matrices $\sigma_{i}$.
We observe that the set of commutation relations for the $SL(2,C)$
algebra is not invariant under the sign change of the rotation
generators, but remains invariant under the sign change of the
boost generators.  Thus, the first and the second solution of this set
of commutation relation consists of
\begin{equation}
J_{i} = {1\over2}\sigma_{i}, \qquad K_{i} = {i\over2}\sigma_{i},
\end{equation}
and
\begin{equation}
J_{i} = {1\over2}\sigma_{i}, \qquad \dot K_{i} = -{i\over2}\sigma_{i},
\end{equation}
respectively.  We call these two representations ``undotted'' and ``dotted''
representations respectively.

When we refer to the $SU(2)$ subgroup of $SL(2,C)$, it is usually understood
to be the subgroup in the undotted representation.  The representation space
consists of
\begin{equation}
\a = \pmatrix{1 \cr 0}, \qquad \b = \pmatrix{0 \cr 1} .
\label{us}
\end{equation}
However, if we take into account the boost generators, there are two
different sets of representation spaces.  We use the notation $\ad$
and $\bd$ for the spinors for the dotted representation, and they also
take the form
\begin{equation}
\ad = \pmatrix{1 \cr 0}, \qquad \bd = \pmatrix{0 \cr 1} .
\label{ds}
\end{equation}
However, these two sets of spinors have quite different Lorentz-boost
properties.  There are therefore four independent spinors.  This is why
the Dirac spinor has four components.

For simplicity, we shall consider rotations around and boosts along
the $z$ direction.  The rotation matrix both in the dotted and
undotted representation is
\begin{equation}
R(\theta) = \exp{\left(-i\theta J_{3}\right)}
= \pmatrix{e^{-i\theta/2} & 0 \cr 0 & e^{i\theta/2}}.
\end{equation}
The boost matrix in the undotted representation is
\begin{equation}
B(\eta) = \exp{\left(-i\eta K_{3}\right)}
= \pmatrix{e^{\eta/2} & 0 \cr 0 & e^{-\eta/2}} ,
\label{boostm}
\end{equation}
while it becomes
\begin{equation}
\dot B(\eta) = \exp{\left(-i\eta \dot K_{3}\right)}
= \pmatrix{e^{-\eta/2} & 0 \cr 0 & e^{\eta/2}} ,
\end{equation}
for the dotted representation, where $K_{3}$ and $\dot K_{3}$
take the form $(i/2)\sigma_{3}$
and $(-i/2)\sigma_{3}$ respectively.  Therefore we have
\begin{equation}
\begin{array}{ll}
B(\eta) \a=e^{\eta/2} \a , & \qquad B(\eta) \b=e^{-\eta/2} \b,  \\
\dot B(\eta) \ad=e^{-\eta/2} \ad , & \qquad \dot B(\eta)\bd=e^{\eta/2} \bd.
\end{array}
\end{equation}

Since the spinorial space is generated by four independent spinors,
the little groups for these spinors are to be constructed accordingly.
As was emphasized in Sec. \ref{sec2}, the generators of the $E(2)$-like
little group are obtained from the contraction procedure defined in
(\ref{n1n2}).  During this process, $J_{3}$ remains unchanged, and
$N_{1}$ and $N_{2}$ applicable to $\a$ and $\b$ become
\begin{equation}
N_{1}=\left(
\begin{array}{cc}
0 & i  \\
0 & 0
\end{array}
\right ), ~~~~~
N_{2}=
\left(
\begin{array}{cc}
0 & 1 \\
0 & 0
\end{array}
\right )
\label{lg1}
\end{equation}
while for $\ad$ and $\bd$ one has to employ
\begin{equation}
\dot N_{1}=\left(
\begin{array}{cc}
0 & 0  \\
-i & 0
\end{array}
\right ), ~~~~~
\dot N_{2}=
\left(
\begin{array}{cc}
0 & 0 \\
1 & 0
\end{array}
\right ).
\label{lg2}
\end{equation}
Related to these generators the transformation matrices are
obtained by $D(u,v)=\exp(-iuN_{1}-ivN_{2})$ and
$\dot D(u,v)=\exp(-iu\dot N_{1}-iv\dot N_{2})$, and explicitly
written as
\begin{equation}
D(u,v)=\left(
\begin{array}{cc}
1 & u-iv  \\
0 & 1
\end{array}
\right ), ~~~~~
\dot D(u,v) =
\left(
\begin{array}{cc}
1 & 0 \\
-u-iv & 1
\end{array}
\right ).
\label{lg3}
\end{equation}
Two of the spinors are invariant under these transformations
\begin{equation}
D(u,v)\a=\a,  \qquad \dot D(u,v)\bd=\bd
\label{gis}
\end{equation}
while the other two are ``gauge-dependent'' in the sense that
\begin{equation}
D(u,v)\b=\b+(u-iv)\a, \qquad  \dot D(u,v)\ad=\ad-(u+iv)\bd. \label{gds}
\end{equation}
The ``gauge-invariant'' spinors of (\ref{gis}) appear as polarized
neutrinos in the real world.  Indeed, it has been revealed that
the polararization of neutrinos is a consequence of gauge
invariance~\cite{hks82}.  There are no neutrinos which correspond
to the gauge-dependent spinors $\ad$ and $\b$, since they represent
unobservable quantities.  It has been shown that these gauge-dependent
spinors are responsible for the gauge dependence of the electromagnetic
four-vector~\cite{hks86}.  This issue will further be elaborated in
Sec.~\ref{massless}.

Let us now construct spin-1 and spin-0 state vectors using two spinors.
In the $SU(2)$ and $O(3)$ regime, we construct the spin-1 and spin-0
wave functions by making the following combinations:
\begin{equation}\label{utensor}
\a\a,\qquad \frac{1}{\sqrt{2}}(\a\b+\b\a),
\qquad\b\b, \qquad \frac{1}{\sqrt{2}}(\a\b-\b\a).
\label{sc1}
\end{equation}
This will be all, if we restrict ourselves to the undotted
representation.  If we restrict ourselves to the dotted representation
we have
\begin{equation}\label{dtensor}
\ad\ad, \qquad\frac{1}{\sqrt{2}}(\ad\bd+\bd\ad),
\qquad \bd\bd, \qquad \frac{1}{\sqrt{2}}(\ad\bd-\bd\ad).
\end{equation}
Here the story is the same for $O(3)$.  It is also possible to make mixed
combinations:
\begin{equation}\label{ud}
\begin{array}{l}
\a\ad,\qquad\frac{1}{\sqrt{2}}(\a\bd+\b\ad),
\qquad \b\bd, \qquad \frac{1}{\sqrt{2}}(\a\bd-\b\ad), \\
\ad\a,\qquad\frac{1}{\sqrt{2}}(\ad\b+\bd\a),
\qquad \bd\b, \qquad \frac{1}{\sqrt{2}}(\ad\b-\bd\a).  \\
\end{array} \label{sc2}
\end{equation}
All of the above four combinations have the same property under
rotations.  However, they become different if the Lorentz boosts
are taken into account.

It is known that the combinations in (\ref{ud}) have the
same transformation property as that of a four-vector under Lorentz
transformations.  It has also been mentioned that particular combintations
in (\ref{utensor}) and ({\ref{dtensor}) behave like
tensors~\cite{berest89,knp86}.  In the next section, we prove that
this is indeed so.  Furthermore, we consider the infinite-momentum/zero-mass
limits of these combinations.  We shall show that the four-vector becomes like
the four-potential, and the tensor becomes like the electromagnetic
fields.

\section{Four-vectors and Tensors from SL(2,C) Spinors}\label{4vec}
\subsection{Four-vectors} \label{massless}
Let us consider the direct products of one dotted and one undotted
spinor as~ $ \a\ad,~ \a\bd,~ \ad\b,~ \b\bd~ $ and a four-vector
$ V^{\mu}=(V_{x},V_{y},V_{z},V_{t}) $ representing a massive particle.
This vector can be expressed in the form of a two-by-two Hermitian matrix
by $V=\sum \sigma_{\mu}V^{\mu}$ or equally well by
$\dot V=\sigma_{0}V_{t}-\sum \sigma_{i}V^{i}$, where $\sigma_{0}$
is the identity matrix.  As the notation suggests $\dot V$ admits
$\dot K_{i}$ as the generators of boosts.  The matrix $V$ explicitly
takes the form
\begin{equation}
V=\left(
\begin{array}{cc}
V_{t}+V_{z} & V_{x}-iV_{y} \\
V_{x}+iV_{y} & V_{t}-V_{z}
\end{array}  \label{matV}
\right),
\end{equation}
and the transformation rules under rotations and boosts are explained in
App. A.  The transformation properties of the direct products of spinors
are derivable from Sec. III.

\begin{table}[h]
\caption{Spinorial combinations for four-vectors}\label{tab1}
\begin{tabular}{@{\hspace{.1in}}cccc@{\hspace{.1in}}}
\hline
Spinorial combinations & Transform like
                     & Under $R(\theta)$ & Under $B(\eta)$ \\
\hline
\hline
$-\a\ad  $  & $ V_{x}-iV_{y} $ & $ e^{-i\theta} $  & $ e^{0}     $  \\
$ \a\bd  $  & $ V_{t}+V_{z}  $ & $ e^{0}        $  & $ e^{\eta}  $ \\
$ -\ad\b $  & $ V_{t}-V_{z}  $ & $ e^{0}        $  & $ e^{-\eta} $ \\
$ \b\bd  $  & $ V_{x}+iV_{y} $ & $ e^{i\theta}  $  & $ e^{0}     $ \\
\hline
\end{tabular}
\end{table}

In view of Table I, we can make
the following identifications between the components $V^{\mu}$
and the direct products of spinors as:
\begin{equation}
\begin{array}{ll}
V_{x}\simeq\frac{1}{2}(\b\bd-\a\ad), & \qquad
V_{z}\simeq\frac{1}{2}(\a\bd+\ad\b), \\
V_{y}\simeq\frac{-i}{2}(\b\bd+\a\ad), & \qquad
V_{t}\simeq\frac{1}{2}(\a\bd-\ad\b).
\end{array}   \label{fv}
\end{equation}
We observe that $V_{x}$ and $V_{y}$ are invariant quantities
considering both rotations around and boosts along the $z$-axis.
On the other hand the components $V_{z}$ and $V_{t}$ are
squeezed, since they transform as
\begin{equation} \label{tfv}
V'_{z}\simeq\frac{1}{2}(e^{\eta}\a\bd+e^{-\eta}\ad\b), \qquad
V'_{t}\simeq\frac{1}{2}(e^{\eta}\a\bd-e^{-\eta}\ad\b)
\end{equation}
under boosts.

\subsection{Scalars and $2^{nd}$-Rank Tensors }
As it has been possible to construct four-vectors
from the direct product of two spinors, it is also possible
to construct scalars and antisymmetric $2^{nd}$-rank tensors.
Antisymmetric combinations $\frac{1}{\sqrt{2}}(\a\b-\b\a)$ and
$\frac{1}{\sqrt{2}}(\ad\bd-\bd\ad)$
are invariant under rotations and boosts, and therefore have
similar transformation properties with that of scalars.

A way to achieve an antisymmetric 2$^{nd}$-rank tensor is
to consider the direct products of two dotted and the direct
products of two undotted spinors
$$ (i) ~~\a\a,~~\a\b,~~ \b\b  \qquad \mbox{and} \qquad
(ii)~~ \ad\ad,~~ \ad\bd,~~\bd\bd.  $$
Under the action of the Lorentz group, particular combinations of
the two categories $(i)$ and $(ii)$ are transformed into combinations of
each other.  Consider an antisymmetric four-by-four matrix
\begin{equation}
T=\left(
\begin{array}{cccc}
0  & -g_{z} & g_{y}  & f_{x} \\
g_{z}  & 0 & -g_{x} &  f_{y} \\
-g_{y} & g_{x} & 0  &  f_{z}  \\
-f_{x} & -f_{y}  & -f_{z} &   0
\end{array} \label{T}
\right ).
\end{equation}
It is well known that $f=(f_{x},f_{y},f_{z})$ and
$g=(g_{x},g_{y},g_{z})$ transform like three-vectors under
rotations.  Transformation properties under boosts are also
well known.  Here we are interested in studying the transformation
properties of the above tensor in terms of the $SL(2,C)$ spinors.
For this purpose, we should be able to write each element of the
tensor $T$ in terms of $SL(2,C)$ spinors.
The boost matrix for the tensor $T$ is
\begin{equation}
B(\eta)=\left(
\begin{array}{cccc}
1  & 0 & 0 & 0 \\
0  & 1  & 0  & 0 \\
0  & 0 & \cosh \eta & \sinh \eta \\
0  & 0 & \sinh \eta & \cosh \eta
\end{array}
\right).
\end{equation}
We note first that
$f_{z}$ and $g_{z}$ are invariant under this boost.  In addition,
these quantities are invariant under the rotation around the $z$-axis,
where the rotation matrix is
\begin{equation}
R(\theta)=\left(
\begin{array}{cccc}
\cos \theta & -\sin \theta & 0 & 0 \\
\sin \theta & \cos \theta & 0  & 0 \\
0 & 0 & 1 & 0 \\
0 & 0 & 0 & 1
\end{array}
\right) .
\end{equation}
The spinor combinations which satisfy these invariance requirements are
\begin{equation}
\frac{1}{\sqrt{2}}(\a\b+\b\a),  \qquad \frac{1}{\sqrt{2}}(\ad\bd+\bd\ad) .
\end{equation}
For this reason, we are allowed to identify them with $f_{z}$ and
$g_{z}$ respectively.  Let us introduce the subsequent notations:
\begin{equation}
\begin{array}{ll}
\kappa_{+}=(f_{x}+ig_{x})+i(f_{y}+ig_{y}), &
\qquad \kappa_{-}=(f_{x}+ig_{x})-i(f_{y}+ig_{y}), \\
\lambda_{+}=(f_{x}-ig_{x})+i(f_{y}-ig_{y}), &
\qquad \lambda_{-}=(f_{x}-ig_{x})-i(f_{y}-ig_{y}).
\end{array} \label{eps}
\end{equation}
We observe that these quantities can be identified with
\begin{equation}
\kappa_{-} \simeq \a\a, \qquad  \kappa_{+} \simeq \b\b, \qquad
\lambda_{-} \simeq \ad\ad,  \qquad  \lambda_{+} \simeq \bd\bd ,
\end{equation}
since they exhibit the same transformation properties which can
directly be derivable from those of
the $SL(2,C)$ spinors given in Sec. \ref{spinors}.

\begin{table}[h]
\caption{Spinorial combinations and $2^{nd}$-rank tensors}\label{tab2}
\begin{tabular}{@{\hspace{.1in}}cccc@{\hspace{.1in}}}
\hline
Spinorial combinations & Transform like
                             & Under $R(\theta)$ & Under $B(\eta)$ \\
\hline
\hline
$\a\a                               $ & $ \kappa_{-}
$ & $ e^{-i\theta}                  $ & $ e^{\eta}  $ \\
$\frac{1}{\sqrt{2}}(\a\b+\b\a)      $ & $ f_{z}
$ & $ e^{0}                         $ & $ e^{0}     $ \\
$\b\b                               $ & $ \kappa_{+}
$ & $ e^{i\theta}                   $ & $ e^{-\eta} $ \\
$\ad\ad                             $ & $ \lambda_{-}
$ & $ e^{-i\theta}                  $ & $ e^{-\eta} $ \\
$\frac{1}{\sqrt{2}}(\ad\bd+\bd\ad)  $ & $ g_{z}
$ & $ e^{0}                         $ & $ e^{0}     $ \\
$\bd\bd                             $ & $ \lambda_{+}
$ & $ e^{i\theta}  $ & $ e^{\eta}   $ \\
\hline
\end{tabular}
\end{table}

In view of Table II, $f$ and $g$ can be
written in terms of $\kappa_{\pm}$ and $\lambda_{\pm}$ as
\begin{equation}
\left(
\begin{array}{c}
f_{x}\\f_{y}\\g_{x} \\g_{y}
\end{array}
\right)
=\frac{1}{4}
\left(
\begin{array}{cccc}
1 & 1 & 1 & 1 \\
-i & i & -i & i \\
-i & -i & i & i \\
-1 & 1 & 1  & -1
\end{array}
\right)
\left(
\begin{array}{llll}
\kappa_{+}\\
\kappa_{-} \\
\lambda_{+} \\
\lambda_{-}
\end{array}
\right),
\end{equation}
or more explicitly as
\begin{equation}
\begin{array}{ll}
f_{x}\simeq\frac{1}{4}(\a\a+\bd\bd+\b\b+\ad\ad), &
\qquad g_{x}\simeq\frac{i}{4}(-\a\a+\bd\bd-\b\b+\ad\ad), \\
f_{y}\simeq\frac{i}{4}(\a\a-\bd\bd-\b\b+\ad\ad), &
\qquad g_{y}\simeq\frac{1}{4}(\a\a+\bd\bd-\b\b-\ad\ad) .
\end{array}  \label{fg}
\end{equation}
Under the Lorentz boost along the $z$-direction, these quantities become
\begin{equation}
\begin{array}{ll}
f'_{x}\simeq\frac{1}{4}[e^{\eta}(\a\a+\bd\bd)+e^{-\eta}(\b\b+\ad\ad)], &
\qquad g'_{x}\simeq\frac{i}{4}[e^{\eta}(\a\a+\bd\bd)
                                         +e^{-\eta}(-\b\b+\ad\ad)], \\
f'_{y}\simeq\frac{i}{4}[e^{\eta}(\a\a-\bd\bd)+e^{-\eta}(-\b\b+\ad\ad)], &
\qquad g'_{y}\simeq\frac{1}{4}[e^{\eta}(\a\a+\bd\bd)+e^{-\eta}(-\b\b-\ad\ad)] .
\end{array}  \label{tfg}
\end{equation}

\section{Massless Particles}\label{masless}
As was shown by In\"on\"u and Wigner~\cite{inonu53}, the group $O(3)$ can
be contracted to $E(2)$, and the contraction procedure is well defined.
It is essentially a large-radius approximation of a finite area on a
spherical surface.  Likewise, contraction of the $O(3)$-like little
group to the $E(2)$-like little group can be defined.  The
infinite-momentum/zero-mass limit corresponds to the large-radius
limit.  Indeed, this limiting process in our case is to let the boost
parameter $\eta$ go to infinity, renormalize the larger components
and let the smaller components go to zero.

In Sec.~\ref{spinors}, we discussed the contraction of the rotation
generators in the $SL(2,C)$ framework.  In this section, we shall study
the infinite-momentum/zero-mass limits of the spinorial combinations
discussed in the preceding section.  This problem has been studied for
four-vectors in the literature.  In this paper, we shall use the
two-by-two representation of the four-vector to study this problem.
As for the tensors, we are dealing with a new problem in this paper.
We shall demonstrate in this section the tensor in (\ref{T})
becomes the Maxwell field tensor for massless particles with spin 1.

\subsection{Four-Potential}
For the case of a massless particle let us consider a photon
moving in the $z$ direction.  The four-momentum vector is then
$p^{\mu}=(0,0,p_{3},p_{0})$.
The mass-shell requirement
$p^{\mu}p_{\mu}=0$, together with the Lorentz condition
 \begin{equation}
\partial_{\mu}A^{\mu}=P_{\mu}A^{\mu}=0.
\label{lg}
\end{equation}
yields
\begin{equation}
A_{t}=A_{z}.
\label{tez}
\end{equation}
In the infinite-$\eta$ limit the identifications (\ref{fv}) represent the
massless case and therefore are rewritten as the components of the
four-potential
\begin{equation}
A_{x}\simeq\frac{1}{2}(\a\ad+\b\bd),~~
A_{y}\simeq\frac{i}{2}(\a\ad-\b\bd),~~
A_{z}=A_{t}\simeq\frac{1}{2}\a\bd.
\end{equation}
In the light cone coordinate system the two-by-two matrix for
the potential takes the form
\begin{equation}
A=\left(
\begin{array}{cc}
A_{u} & A_{x}-iA_{y} \\
A_{x}+iA_{y} & 0
\end{array}
\right),
\end{equation}
where $A_{u}=(A_{t}+A_{z})/\sqrt{2}$ and $A_{v}=(A_{t}-A_{z})/\sqrt{2}=0$.
The vanishing of $A_{v}$ in the infinite-$\eta$ limit is in accordance
with the gauge condition (\ref{lg}).  The transformation properties
of the four-vector potential under the action of the little group
represented by a four-by-four matrix is readily available in the
literature~\cite{kiwi87jm}.  Now, we have a two-by-two matrix
for the four-potential, and the $E(2)$-like little group acts on
$A$ as
\begin{equation}
A'=D(u,v)~A~D(u,v)^{\dagger},
\end{equation}
where $D(u,v)$ is given in (\ref{lg3}).  Then
\begin{equation}
A'=\left(
\begin{array}{cc}
A_{u}+uA_{x}+vA_{y} & A_{x}-iA_{y} \\
A_{x}+iA_{y} & 0
\end{array}
\right).
\end{equation}

We noted that, for massless particles, the spinors of  (\ref{gis})
are gauge-invariant, while those of  (\ref{gds}) are gauge-dependent.
The gauge-dependence of the four-potential comes from those
gauge-dependent spinors.  This is also why the four-potential is not
directly observable.  Is it then possible to construct state vectors
solely from gauge-invariant spinors.  This question was first addressed
by Weinberg~\cite{wein64}.  As we shall see in the following section,
the gauge-invariant Maxwell field tensor is composed of those state vectors.

\subsection{Maxwell Field Tensor}
If we follow the contraction procedure, the $f'$ and $g'$components
become the $E$ and $B$ components, where
\begin{equation}
\begin{array}{ll}
E_{x}\simeq\frac{1}{4}(\a\a+\bd\bd), &
\qquad B_{x}\simeq\frac{i}{4}(-\a\a+\bd\bd) \\
E_{y}\simeq\frac{i}{4}(\a\a-\bd\bd), &
\qquad B_{y}\simeq\frac{1}{4}(\a\a+\bd\bd)
\end{array}  \label{EB}
\end{equation}
and hence $E_{x}=B_{y}$ and $E_{y}=-B_{x}$, which manifest
the properties of electromagnetic fields for a circularly
polarized photon propagating in the $z$ direction.

We can now write the tensor as
\begin{equation}
F =\left(
\begin{array}{cccc}
0  & 0 & B_{y}  & E_{x} \\
0  & 0 & -B_{x} &  E_{y} \\
-B_{y} & B_{x} & 0  &  0  \\
-E_{x} & -E_{y}  & 0 &   0
\end{array}
\right ) .
\end{equation}
This expression is only for the field propagating along the $z$
direction.  In general, this tensor can be rotated to suit an arbitrary
direction.  Then, the tensor takes the familiar form
\begin{equation}
F =\left(
\begin{array}{cccc}
0  & -B_{z} & B_{y}  & E_{x} \\
B_{z}  & 0 & -B_{x} &  E_{y} \\
-B_{y} & B_{x} & 0  &  E_{z}  \\
-E_{x} & -E_{y}  & -E_{z} &   0
\end{array}
\right ) .
\end{equation}
It is important to note that the second-rank
tensor $T$ in (\ref{T}) is quite different from the above
expression.  The tensor $T$ is for a massive particle and forms a
representation space for the $O(3)$-like little group.

{}From the $SL(2,C)$ stand point, there is a crucial difference
between the four-potential and the electromagnetic field.
We have seen in the preceding section that the potential is
represented by the direct products of one dotted and one
undotted spinor.  On the other hand, electromagnetic fields are
only represented by the gauge-invariant spinors of (\ref{gis}),
where the direct product is between spinors which are either both
dotted or both undotted.  Furthermore, unlike the transformations
on the four-potentials which can be made unitary~\cite{hks85},
the transformations on the electromagnetic fields are nonunitary.

\section*{Concluding Remarks}
We are not the first to construct all possible representations
using two $SL(2,C)$ spinors.  In this paper, we considered
them in terms of Wigner's concept of little groups for internal space
time symmetries.  We considered them also in terms of the
In\"on\"u-Wigner contraction which leads to the idea of the $E(2)$-like
little group as an infinite-momentum/zero-mass limit of the $O(3)$-like
little group.  We now have achieved a full understanding of the Maxwell
fields and potentials in terms of Winger's little group.

\section*{Acknowledgements}
One of the authors (SB) wishes to thank J. Sucher for making her visit
possible and the Elementary Particle Theory Group in the University of
Maryland for their hospitality.

\begin{appendix}
\section{Two-by-two Matrix Representation of the Four-vector}
The two-by-two matrix representation $V$ for a four-vector as given in
(\ref{matV}) is well known.  It simply works!  In this appendix, we would
like to show why it works.  The rotation subgroup of $SL(2,C)$ is well
known, and is called $SU(2)$.  Then rotation on this matrix is achieved by
\begin{equation}\label{ex}
V' = R V R^{\dagger} .
\end{equation}
The rotation matrices are Hermitian, and it Hermitian conjugate
$R^{\dagger}$ is its inverse.

On the other hand, the story is quite different for boost matrices.
Consider the boost matrix $B_{3}$ along the $z$ direction as given in
(\ref{boostm}).  The boost along an arbitrary direction takes
the form
\begin{equation}
B(\eta) = S B_{3}(\eta) S^{\dagger} ,
\end{equation}
where $S$ is a rotation matrix.  Unlike the rotation matrix, the boost is not
represented by a Hermitian matrix.  The inverse of this matrix is not its
Hermitian conjugate, but it is its ``dot-conjugate'':
\begin{equation}
\dot{B}(\eta) = S B_{3}(-\eta) S^{\dagger} .
\end{equation}
Is it then possible to obtain this form by a matrix conjugation?  Let
us consider the matrix
\begin{equation}
g = i\sigma_{2} = \pmatrix{0 & 1 \cr -1 & 0} .
\end{equation}
This matrix commutes with $\sigma_{2}$ but anticommutes with $\sigma_{1}$
and $\sigma_{3}$.  For this reason,
\begin{equation}
\sigma_{i} = - g \sigma_{i}^{T} g^{-1},
\end{equation}
where $\sigma_{i}^{T}$ is the transpose of $\sigma_{i}$.  Let us call
the above operation the ``g-conjugation''.  The g-conjugation of the
rotation matrix results in its Hermitian conjugate:
\begin{equation}
R^{-1} = R^{\dagger} = g R^{T} g^{-1} , \qquad
S^{-1} = S^{\dagger} = g S^{T} g^{-1} .
\end{equation}
The g-cojugation of the boost matrix results in the inverse or its
dot-conjugation:
\begin{equation}
\dot{B} = B^{-1} = g B^{T} g^{-1} .
\end{equation}

Next, what does the matrix $g$ do on the spinors.  It converts $\alpha$
into $\beta$, and $\beta$ into -$\alpha$.  It converts also
$\dot{\alpha}$ into $\dot{\beta}$, and $\dot{\beta}$ into -$\dot{\alpha}$.
With this point in mind, we can consider the matrix
\begin{equation}
X = \pmatrix{\alpha \dot{\beta} & -\alpha\dot{\alpha} \cr
    \beta \dot{\beta} & -\beta \dot{\alpha}}
   = \pmatrix{\alpha \cr \beta} \pmatrix {\dot{\beta} & -\dot{\alpha}} .
\end{equation}
The rotation of this matrix is done by
\begin{equation}
X' = R X R^{-1} = R X R^{\dagger} ,
\end{equation}
while it is boosted by the formula
\begin{equation}
X' = B X B = B X B^{\dagger} .
\end{equation}

Furthermore, the components of the above $X$ matrix can be identified
with their respective counterparts in the matrix $V$ given in
(\ref{matV}).

\section{Comparison of the Present Notation with the Conventional Notation}
The purpose of this appendix is to clarify the relation
between the present choice of the representation space with the
one that frequently appears in literature~\cite{berest89}.
The dotted and the undotted spinors are
denoted by $\xi^{a}=(\xi^{1},\xi^{2})$ and
$\eta^{\dot a}=(\eta^{\dot 1},\eta^{\dot 2}) $, where
the action of any element of the group $SL(2,C)$ transforms
the undotted spinors as:
\begin{equation}
\xi'^{1}=a \xi^{1}+c \xi^{2}, \qquad  \xi'^{2}=b \xi^{1}+d \xi^{2}.
\end{equation}
Here $a, b, c, d$ are complex functions of the six parameters
of the group and are subject to $ad-bc=1$.  As to the transformation
properties of the dotted spinors one has to introduce the
complex conjugate transformation:
\begin{equation}
\eta'^{\dot 1}=a^{\ast} \eta^{\dot 1}+c^{\ast} \eta^{\dot 2},
\qquad  \eta'^{\dot 2}=b^{\ast} \eta^{\dot 1}+d^{\ast} \eta^{\dot 2}.
\end{equation}
This is possible if one sets $\xi^{1}=(1,0)^{T},~\xi^{2}=(0,1)^{T}$
but $\eta^{\dot 1}=(0,1)^{T},~\eta^{\dot 2}=(1,0)^{T}$.
By comparison we thus have $\xi^{a}=(\a, \b)$ and
$\eta^{\dot a}=(\bd, -\ad)$.  Since choosing the above
representation space admits only one of the solutions
(i.e., solution (2.3)) of the commutation relations of the
Lie algebra of $SL(2,C)$, the representation space as given in
(\ref{us}) and (\ref{ds}) is actually a natural
consequence of having two nonequivalent solutions to the
commutation relations of the algebra rather than choosing a
convention.
\end{appendix}

\end{document}